\documentclass[12pt,preprint]{aastex}

\shorttitle{Horizontal Branch stars in $\omega$\,Centauri}
\shortauthors{Moni Bidin et al.}

\begin{document}

\title{The peculiar properties of horizontal branch stars in $\omega$\,Centauri
\thanks{Based on observations with the ESO Very Large Telescope at Paranal Observatory, Chile (proposal ID 076.D-0810)}}

\author{C. Moni Bidin and S. Villanova}
\affil{Departamento de Astronom\'ia, Universidad de Concepci\'on, Casilla 160-C, Concepci\'on, Chile;}
\email{cmbidin@astro-udec.cl}
\author{G. Piotto}
\affil{Dipartimento di Astronomia, Universit\`a di Padova, Vicolo dell'Osservatorio 3, I-35122 Padua, Italy}
\and
\author{S. Moehler}
\affil{European Southern Observatory, Karl-Schwarzschild-Str. 2, 85784 Garching, Germany}
\and
\author{F. D'Antona}
\affil{INAF, Osservatorio Astronomico di Roma, Via Frascati 33, 00040 Monteporzio Catone, Roma, Italy}

\begin{abstract}
We measured temperatures, gravities, and masses for a large sample of blue horizontal branch stars in
$\omega$\,Centauri, comparing the results with theoretical expectations for canonical and He-enriched stars, and with
previous measurements in three other clusters. The measured gravities of $\omega$\,Cen stars are systematically lower
than canonical models, in agreement with expectations for He-enriched stars, and contrary to that observed in the
comparison clusters. However, the derived masses are unrealistically too low as well. This cannot be explained by low
gravities alone, nor by any of the other parameters entering in the calculation. We find that the same stars are not
brighter than their analogs in the other clusters, contrary to the expectations of the He-enrichment scenario.
The interpretation of the results is thus not straightforward, but they reveal an intrinsic, physical difference
between HB stars in $\omega$\,Cen and in the three comparison clusters.
\end{abstract}


%%%%%%%%%%%%%%%%%%%%%%%%%%%%%%%%%%%%%%%%%
%%%%%%%%%%%%%%%%%%%%%%%%%%%%%%%%%%%%%%%%%

\section{INTRODUCTION}
\label{s_intro}

The nature of the second parameter, aside from metallicity, which determines the morphology of the horizontal branch (HB)
in globular clusters (GCs), is one of the most longstanding problems of modern astrophysics. In fact, a lower
metallicity favors the formation of hotter and bluer HB stars, but clusters with the same metallicity can
show very different HB morphology \citep{Sandage67}, and an HB extended far toward the blue is observed
even in some metal-rich GCs \citep[e.g.,][]{Rich97}. The helium abundance has been early proposed, among others,
as this second parameter (\citealt{Sweigart97,DAntona02}; see \citealt{Catelan09b}, for a review) because, during the
He-burning phase, helium-rich stars are expected to be hotter than objects of canonical composition. This model has
recently drawn much attention, triggered by the discovery of multiple stellar populations in GCs
\citep{Piotto05,Piotto07}. In fact, \citet{Piotto05} showed that a different metallicity is not the cause of the
main-sequence split observed in $\omega$\,Centauri \citep{Bedin04}, and the only explanation is that the bluer sequence
is greatly enriched in helium, about 50\% more He-rich (Y=0.38) than in normal metal-poor GC stars
\citep{Norris04,Piotto05}. In this scenario, the blue HB stars observed in many GCs could be the progeny of
the He-enriched second stellar generation. Unfortunately, diffusion processes completely alter the surface
chemical abundances of hot HB stars \citep[e.g.,][]{Behr99}, preventing a direct demonstration of the
connection between multiple populations and HB morphology. Nevertheless, an increased helium content can
be indirectly deduced from other observable quantities, because He-enriched HB stars are predicted to be
brighter \citep{Sweigart87} and to occupy different loci in the temperature--gravity plane
\citep{Moehler03}.

In this Letter, we present the results of our investigation aimed to search for an indirect indication of
helium enrichment among blue HB stars in $\omega$\,Centauri, to test the He-enrichment scenario and its
predicted effects on the HB morphology. This cluster is the ideal target for our purpose, because
it hosts a very complex stellar population, comprising three known MSs and six sub-giant branches
\citep{Bellini10}.

%%%%%%%%%%%%%%%%%%%%%%%%%%%%%%%%%%%%%%%%%
%%%%%%%%%%%%%%%%%%%%%%%%%%%%%%%%%%%%%%%%%

\section{OBSERVATIONS AND DATA ANALYSIS}
\label{s_obs}

\begin{figure}
\epsscale{1.}
\plotone{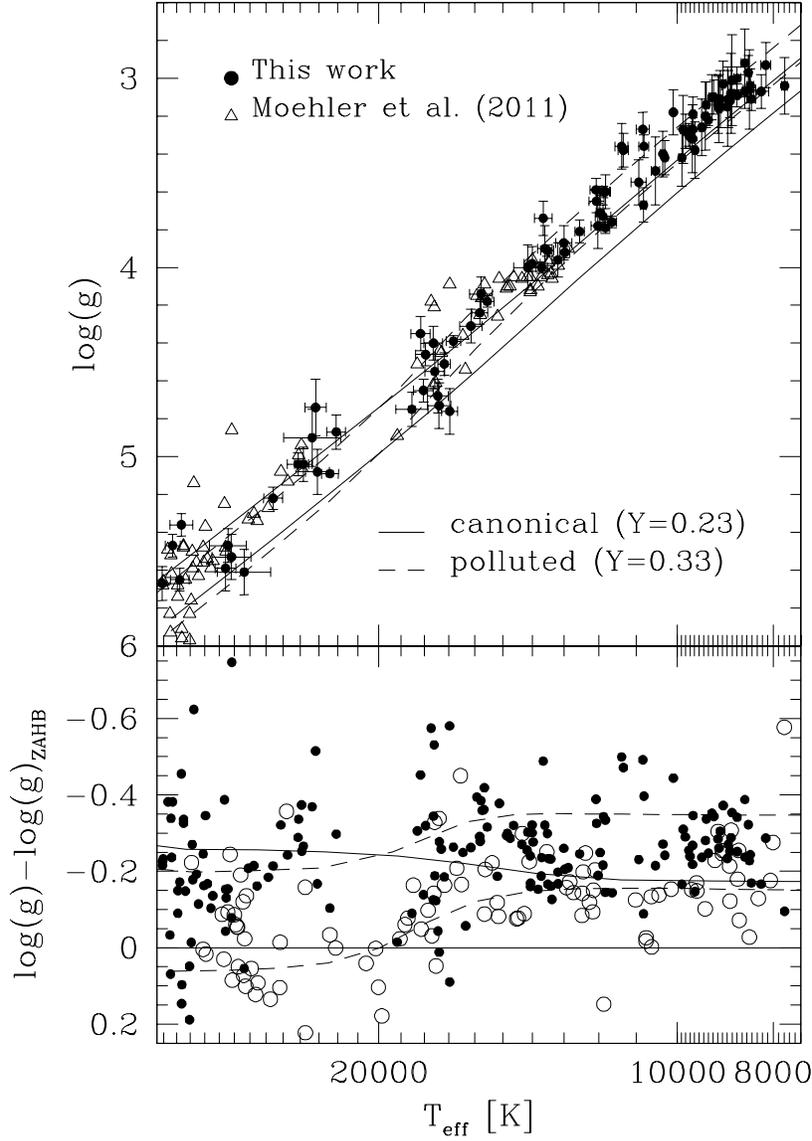}
\caption{{\it Upper panel}: distribution of $\omega$\,Cen stars in the temperature-gravity plane. The
Zero-Age and Terminal-Age HB (ZAHB and TAHB, respectively), for both canonical and He-enriched models are
also indicated. {\it Lower panel}: comparison of stars in $\omega$\,Cen (full dots)
and members of three other clusters (open circles). The vertical coordinate is the difference between the
stellar gravity and the corresponding value of the canonical ZAHB at the same temperature. The plot is thus
analogous to the T$_\mathrm{eff}$--$\log{\mathrm (g)}$ space of the upper panel, but the horizontal axis
coincides with the canonical ZAHB.}
\label{f_tg}
\end{figure}

We selected 116 target stars from the ground-based photometry of \citet{Bellini09}.
They span a wide portion of the cluster HB, from the blue edge of the RR Lyrae gap to the Blue
Hook objects (T$_\mathrm{eff}\geq$ 33\,000~K) 5 mag fainter. In this Letter, we will
focus on the comparison of $\omega$\,Cen stars with other clusters and theoretical
models. We therefore limit the analysis to T$_\mathrm{eff}\leq$33\,000~K,
because hotter Blue Hook stars are not included in the canonical models and are not present in
the comparison clusters. A full analysis of the results, including the Blue Hook, will be
presented in a forthcoming paper.

The data were collected at Paranal Observatory in service mode between 2006 January and April,
with FORS2@UT1 in MXU mode. The selected 600B grism, coupled with 0$\farcs$5-wide slits,
gave spectra of resolution R$\approx1600$ from the atmospheric cutoff to approximately
5900\,\AA. Three 45 minute spectra for faint stars and two 45 minute spectra for bright ones were
acquired. Data were reduced with the FORS
pipeline\footnote{\small{www.eso.org/sci/data-processing/software/pipelines/index.html}}, and the
spectra were extracted under IRAF\footnote{\small{IRAF is distributed by the National Optical
Astronomy Observatories, which is operated by the Association of Universities for Research in
Astronomy, Inc., under cooperative agreement with the National Science Foundation.}}, subtracting
the nearby sky spectrum within the same slit. Finally, the spectrum of the standard star LTT4816
\citep{Hamuy92}, secured during observations, was used to flux-calibrate the object spectra, whose
resulting signal-to-noise ratio was between 50 and 150. Heliocentric radial velocities (RV) were measured
with the IRAF {\it fxcor} task, cross-correlating \citep{Tonry79} the spectra with synthetic templates of
adequate parameters as estimated from the stellar position in the color-magnitude diagram (CMD). Considering
the internal velocity dispersion of the cluster ($\sim$13\,km\,s$^{-1}$, \citealt{Sollima05}) and the errors
of measurements (about 30~km~s$^{-1}$), the RV of all the observed stars is consistent with cluster
membership.

The atmospheric parameters of target stars were measured fitting the observed Balmer and helium
lines with stellar model atmospheres, computed with ATLAS9 \citep{Kurucz93}. We used Lemke's
version\footnote{\small{For a description see
http://a400.sternwarte.uni-erlangen.de/$\sim$ai26/linfit/linfor.html}} of the LINFOR program (developed
originally by Holweger, Steffen, and Steenbock at Kiel University) to compute a grid of synthetic
spectra. Stars showing iron lines in the 4450--4600 \AA\ region, indicating active atmospheric
diffusion \citep{Moehler99}, or being hotter than 13\,000~K (as
deduced from the position in the CMD), were fitted with metal-rich models ([M/H]=+0.5) with varying
surface helium abundance, to account for the effects of radiative levitation of heavy elements
\citep[e.g.,][]{Moehler00}. This was done even for five stars hotter than 11\,500~K not satisfying
these criteria, because the observed helium lines were clearly too weak compared to the fitted models
when diffusion was not taken into account. The cooler stars were fitted with cluster-metallicity models
([M/H]=$-$1.5) with helium abundance fixed at the solar value, because their He lines are very weak and not
observed at our resolution. In a few cases of doubt about the correct model to use, we adopted the set of
model spectra that returned the lower $\chi^2$ of the fit. The best fit to the observed spectra was
obtained by means of the routines developed by \citet{Bergeron92} and \citet{Saffer94}, as modified by
\citet{Napiwotzki99}, which employ a $\chi^2$ test. The spectral lines used in the procedure included
the Balmer series from H$_\beta$ to H$_{12}$, except for H$_\epsilon$ to avoid the blended
Ca\,II~H line, four He\,I lines (4026 \AA, 4388 \AA, 4471 \AA, and 4921 \AA) for stars whose
helium abundance was not kept fixed, and the He\,II lines 4542 \AA\ and 4686 \AA\ when visible
in the spectra of the hottest stars.
The routines estimate the errors on the derived parameters from the $\chi^2$ of the fit
\citep[see][]{Moehler99}, but neglect all other sources of errors (e.g., defects in normalization,
flat-field correction, sky subtraction). We therefore obtained a better estimate of the true errors
multiplying the output values by 3 (R. Napiwotzki, private communication).

Stellar masses were calculated from the measured temperatures and gravities, through the relation:
\begin{equation}
\log{\frac{M}{M_{\sun}}}=\log{\frac{g}{g_{\sun}}}-4\cdot \log{\frac{T}{T_{\sun}}}+\log{\frac{L}{L_{\sun}}},
\label{eqmass1}
\end{equation}
where
\begin{equation}
\log{\frac{L}{L_{\sun}}}=-0.4\cdot(V - (m-M)_0 - 3.1\cdot E(B-V) + BC - 4.74).
\label{eqmass2}
\end{equation}
We assumed T$_{\sun}$=5777 K, $\log{\mathrm{g}_{\sun}}$=4.44,
(m-M)$_0$=13.75$\pm$0.13 \citep{Vandeven06}, and E($B-V$)=0.12$\pm$0.01 \citep[][2010 December Web
version]{Harris96}. The bolometric correction (BC) was derived from effective temperatures through the empirical
calibration of \citet{Flower96}. Errors on masses were derived from propagation of errors.

%%%%%%%%%%%%%%%%%%%%%%%%%%%%%%%%%%%%%%%%%
%%%%%%%%%%%%%%%%%%%%%%%%%%%%%%%%%%%%%%%%%

\section{RESULTS}
\label{s_results}

Our results are plotted in the upper panel of Figure~\ref{f_tg}, where we show the position of the program stars
in the temperature-gravity space, superimposed to the theoretical zero-age and terminal-age HB (ZAHB and TAHB,
respectively) from \citet{Moehler03}, for canonical (Y=0.23) and He-enriched (Y=0.33) models. In the same figure
we include 78 stars from \citet{Moehler11}, who measured the atmospheric parameters with the same procedure
as in the present work, but based on medium-resolution FLAMES spectra, and a different set of model spectra for
stars above 20\,000~K. The two sets of data behave very similarly in the T$_\mathrm{eff}$-$\log{\mathrm (g)}$
plane. The comparison of the eight stars in common confirms the good agreement between the two works: the mean
difference in gravity is null ($\leq0.01$ dex), while the difference in temperature is small (155~K), and
becomes negligible (25~K) after the exclusion of two stars with very large errors ($\geq$1500~K). Our mass
estimates are on average higher by 0.035 M$_\sun$, an offset accounted for by the fainter magnitudes (0.09 mag
the mean difference) from the \citet{Castellani07} catalog used by Moehler et al. Given the excellent agreement,
we will merge the two datasets together.

\begin{figure}
\epsscale{1.05}
\plotone{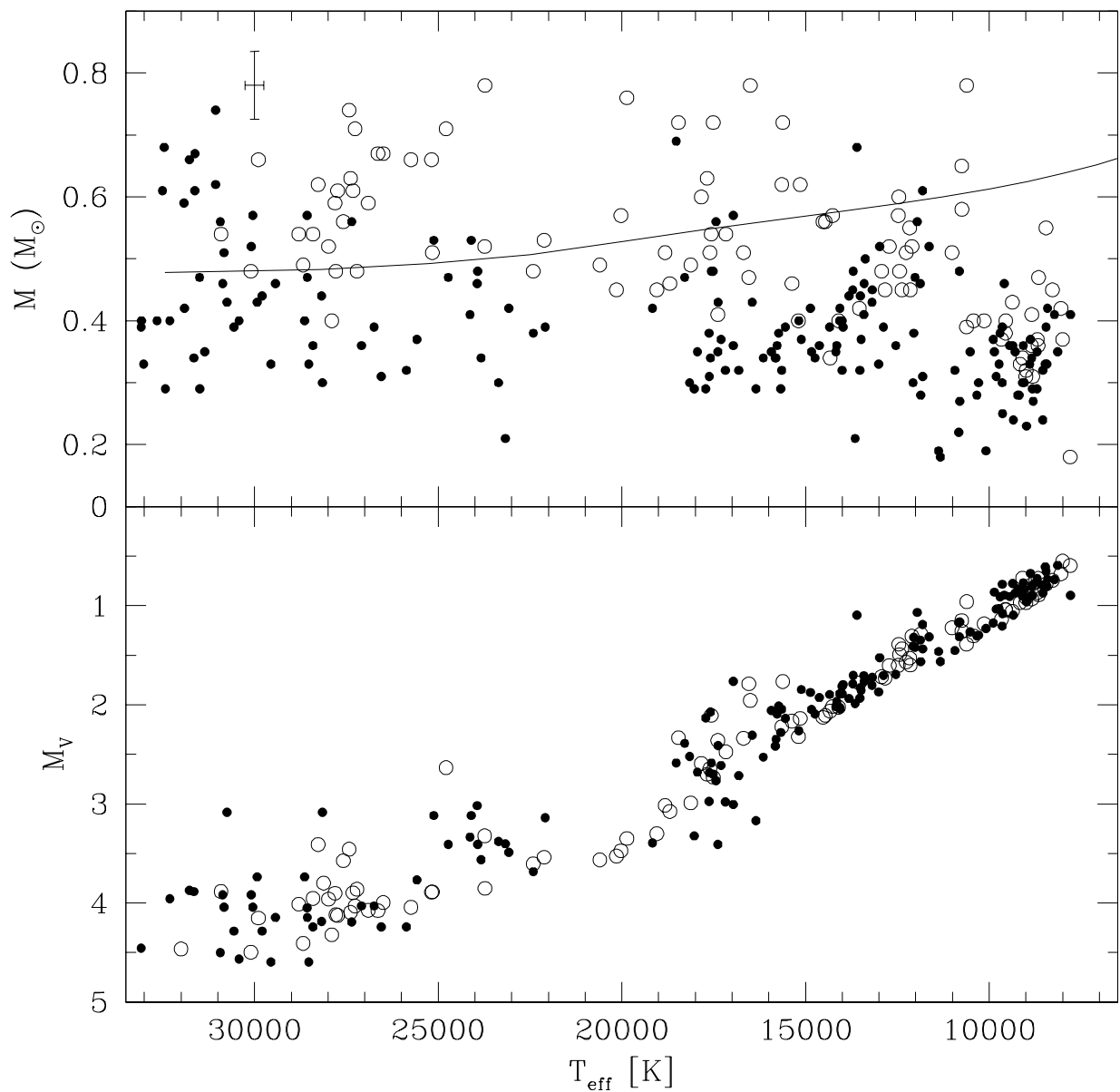}
\caption{{\it Upper panel}: stellar masses, calculated through Equation~(\ref{eqmass1}). The line indicates
the canonical model expectation. {\it Lower panel}: absolute magnitudes, estimated from distance moduli and
reddening given in Section~\ref{s_obs} and \citet{Moni07,Moni09}. Different symbols are used for $\omega$\,Cen
 and other cluster stars, as in the lower panel of Figure~\ref{f_tg}.}
\label{f_masabs}
\end{figure}

The surface gravity of stars cooler than $\sim$18\,000~K is systematically lower by about 0.2-0.3 dex with
respect to canonical models, while they closely follow the trend of He-enhanced models at all temperatures.
In the lower panel of Figure~\ref{f_tg} we compare these results with similar measurements obtained in three GCs,
namely NGC\,6752 \citep{Moni07}, M80 and NGC\,5986 \citep{Moni09}. We adopt as vertical coordinate the
difference between the measured $\log{\mathrm (g)}$ and the value of the canonical ZAHB at the corresponding
temperature. The comparison reveals that the HB stars in $\omega$\,Cen and in the other GCs behave very
differently. We note that the stars of the comparison clusters do not completely agree with
canonical models, whose expectation is to find the majority of the objects next to the ZAHB, and not to the
TAHB as observed. Even so, $\omega$\,Cen stars clearly show
lower gravities -- at a given effective temperature -- with respect to stars in other
GCs. Observational errors tend to mask the general trend, but there is an offset of 0.15 dex at the cooler end,
which decreases at higher temperatures and fades out around 18\,000~K, to reappear even larger ($\geq$0.2 dex)
among hot stars at 25\,000-28\,000~K. However, we find a problem in the estimate of stellar masses, summarized
in the upper panel of Figure~\ref{f_masabs}: while the results in the comparison clusters roughly agree with
expectations \citep[see][for a complete discussion]{Moni07,Moni09}, the masses of $\omega$\,Cen stars are
constantly underestimated at all temperatures. Interestingly enough, \citet{Moehler06} found very similar
results for HB stars in NGC\,6388. This is a very peculiar cluster, as it has an extended, blue HB
\citep{Rich97} despite its high metallicity ([Fe/H]=$-$0.6). This has been interpreted in terms of an extreme
He-enrichment \citep[up to Y=0.40,][]{Caloi07}, as in the case of its
twin cluster, NGC\,6441. While \citet{Moehler06} cast doubts on their results due to the
large uncertainties caused by stellar crowding in this compact, bulge
cluster, these results are very similar to what we find now in $\omega$\,Cen.

%%%%%%%%%%%%%%%%%%%%%%%%%%%%%%%%%%%%%%%%%
%%%%%%%%%%%%%%%%%%%%%%%%%%%%%%%%%%%%%%%%%

\section{DISCUSSION}
\label{s_discussion}

\begin{figure}
\epsscale{0.7}
\plotone{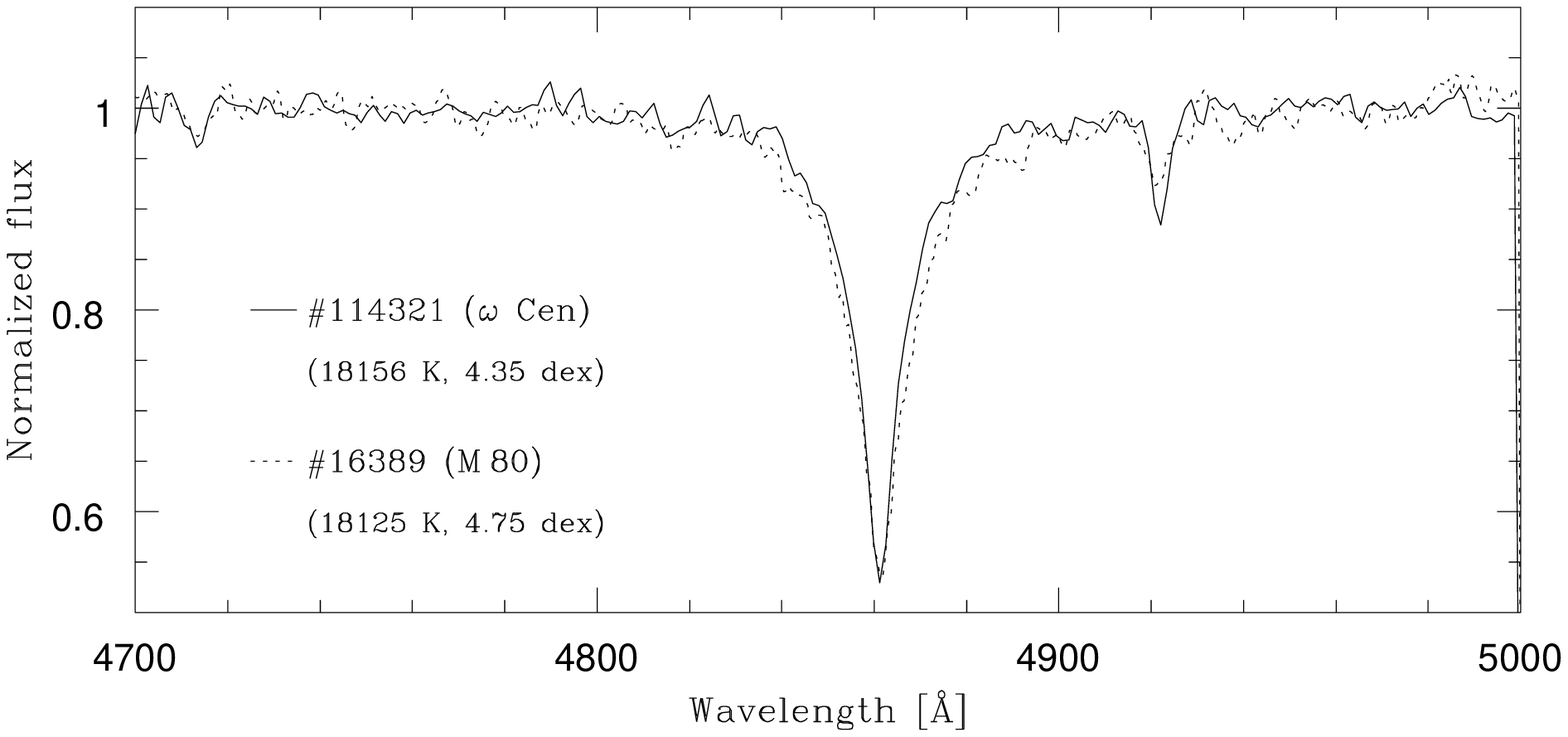}
\caption{Comparison of the H$_\beta$ lines of one of our program star (full line) and a HB
star in M\,80 at the same temperature. Derived temperatures and gravities are given.}
\label{f_spec}
\end{figure}

In the temperature-gravity plane, the $\omega$\,Cen HB stars match the expectations of He-enriched models rather
than canonical ones. However, the resulting
underestimate of their mass prevents us from straightforwardly concluding that this is evidence of helium
enrichment. In fact, the progenies of He-rich stars in the HB phase should not be noticeably less
massive than stars of canonical composition \citep{DAntona04}, and the difference at any temperature is expected
to be tiny \citep[$\leq$0.03~M$_\sun$,][]{Moehler03}. Moreover, the derived masses are on average well below the
value required to ignite helium in the core ($\sim$0.45~M$_\sun$).

The easiest interpretation of our observations is the presence of a systematic error, biasing the results toward
lower gravities and, as a consequence, lower masses. However, in this work we used the same instrument, software,
and models as \citet{Moni07,Moni09}, finding a clear difference between $\omega$\,Cen and the other clusters,
while our measurements well agree with \citet{Moehler11}, who also investigated $\omega$\,Cen but with a
different instrument, higher resolution, different models for stars hotter than 20\,000~K, and only a subset of
Balmer lines. Therefore, even if we cannot completely exclude an observational bias with respect to theoretical
expectations, the difference between $\omega$\,Cen and the three comparison clusters must be real: {\it HB stars
in $\omega$\,Cen are intrinsically different to their analogs in other GCs}. The same conclusion can be drawn
even if the offset is a product of the inadequacy of the employed models: in this case, they would be reproducing
sufficiently well the atmospheric structure of the HB objects in the comparison clusters, but not in $\omega$\,Cen,
hence a physical difference would be present.

The observed trends cannot be due to a wrong (hotter) temperature scale. In fact, for each star we translated
the measured temperature in a reddening estimate, comparing the observed ($B-V$) color to the theoretical value
obtained interpolating the \citet{Kurucz93} grid, for the same metallicities as the model spectra used in the
fits. The average value is E($B-V$)=0.114, in perfect agreement with the literature, and with no significant
trend along the HB. A temperature scale hotter by 10\% (5\%) would have caused an overestimate of reddening by
0.04 (0.02) mag at 10\,000~K. In Figure~\ref{f_spec} we compare the H$_\beta$ line profiles of two stars
with similar temperature in M\,80 and $\omega$\,Cen: the line core and depth are very similar, indicating no
noticeably difference in temperature, but the star in M\,80, whose measured gravity is 0.40~dex higher, shows
wider wings. This comparison indicates that the peculiar stellar gravities reflect a real difference in the
spectra of the target stars. We are aware that some stellar parameters unaccounted for in our study, such as
stellar wind and rotation, can cause wider line wings mimicking a difference in gravity, and this degeneracy
cannot be avoided at our resolution. The underestimate of gravities can also be caused by the stratification of
elemental abundances due to diffusion processes \citep{Leblanc10}. However, a systematic difference in one of
these parameters, affecting $\omega$\,Cen but not the other clusters, would be even more puzzling, and it is
harder to postulate at this stage. It must also be noted that the mass underestimate does not completely follow
the trend of the difference in gravity: while this seems to vary with temperature as noted before, the masses
in $\omega$\,Cen are constantly underestimated by $\sim$0.15 M$_\sun$ for T$_\mathrm{eff}\geq$10\,000~K. For
example, at T$_\mathrm{eff}\sim$16\,000-18\,000~K there is a clear offset in masses, but not in gravities. This
indicates that low gravities may play a role, but at least another effect should be at work, causing the
observed mass underestimate.

All the comparison clusters have similar metallicity \citep[{[Fe/H]}$\approx -$1.5][]{Harris96}, while
$\omega$\,Cen shows a large spread up to [Fe/H]=$-$0.6 \citep[e.g.,][]{Sollima05}. Nevertheless, it is unlikely
that the higher metallicity is the origin of the peculiar results in $\omega$\,Cen: the largest differences are
found for stars hotter than $\sim$11\,500~K, whose surface abundances are altered by diffusion processes.
\citet{Behr03} and \citet{Pace06} showed that, in the presence of diffusion, the surface abundance patterns are
very similar in clusters of very different initial metallicity. The stars in all the clusters should therefore
show the same behavior independently of their primordial metal content, especially at 15\,000-16\,000~K where
diffusion
reaches its maximum strength \citep{Moni09}. Moreover, \citet{Moehler00} found no peculiarity in the measured
gravity and mass of two stars in 47\,Tuc ([Fe/H]=$-$0.7), although using low-metallicity models. We repeated the
measurements assuming different values of the model metallicity, to test how this parameter can affect the
results. We found small differences in the stellar parameters, but the general behavior was unaltered: a higher
model metallicity indeed returned slightly higher gravities, but higher temperatures too. As a consequence, the
points were shifted almost parallel to the theoretical tracks in the T$_\mathrm{eff}$-$\log{\mathrm (g)}$ plane,
while the masses were increased by less than 0.05~M$_\sun$.

The blanketing effect should be lower in metal-poor stars than in the solar-metallicity stars used to calibrate
the adopted T$_\mathrm{eff}$-BC relation, and this could cause the underestimate of the BC and of the mass.
The adopted BC should be a good approximation for the stars hotter than the Grundahl jump \citep{Grundahl99},
because radiative levitation increases their surface abundances to super-solar values. As already noted, the
effect should be independent of their primordial metallicity, thus the BC does not explain the offset of
$\omega$\,Cen with respect to the other GCs. At cooler temperatures, the offset could be
explained by the BC if $\omega$\,Cen stars were more metal-poor than in the other GCs, thus decreasing their BC
by $\sim$0.4 mag. Indeed, $\omega$\,Cen hosts a metal-poor sub-population at
[Fe/H]=$-$2 \citep[e.g.,][]{Pancino11}, but the BC varies by less than 0.15~mag for stars at 8\,000-10\,000~K in
the whole range between solar metallicity and [Fe/H]=$-$2 \citep{Cassisi99,Alonso99}. A wrong distance modulus
or reddening could also cause wrong mass estimates, but the required correction to increase the masses by
0.1~M$_\sun$ is huge ($\geq$0.4 mag in distance modulus, $\geq$0.13 mag in reddening). The recent
literature estimates agree on these quantities within 0.1 and 0.01 mag, respectively, allowing only
negligible variations of the mass estimates. Even an offset in the photometric zero-point could cause
the observed offset, but comparing the $V$ mag of \citet{Bellini09} with the values of
\citet{Castellani07} and \citet{Momany04} we found only a small difference ($\leq$0.1 mag) in the
direction opposite to that required, with the magnitudes used here being brighter than in the other catalogs.
In conclusion, none of the photometric parameters (stellar magnitude, BC, distance modulus, and reddening)
entering in the calculation of the stellar mass through Equation~(\ref{eqmass2}) offers a viable explanation of
our results.

He-enriched HB stars are expected to be brighter than analogous objects of canonical composition
\citep[e.g.,][]{Caloi07,Catelan09}, and the increased luminosity should balance the lower gravities in
Equation~(\ref{eqmass1}), returning a similar mass. All other parameters being the same, M$_\mathrm{V}$ should
be brighter by 0.25-0.38 mag to compensate a decrease of 0.10-0.15 dex in gravity. On the contrary, in
the lower panel of Figure~\ref{f_masabs} the perfect match between the absolute magnitudes of the HB of
$\omega$\,Cen and the other clusters is clear. The too low values obtained for the stellar masses
could therefore also be interpreted as due to the lack of increased flux, instead of too low gravity
estimates. It could be argued that He-enriched stars could not necessarily be brighter in the $V$ band,
because the bolometric luminosity is the quantity entering in Equation~(\ref{eqmass1}). Thus, the detailed
spectral energy distribution (SED) of He-enriched and canonical stars is required to properly deduce their
luminosity from the $V$ magnitude through Equation~(\ref{eqmass2}). However, great differences are not
expected for stars hotter than 12\,000~K, because the diffusion processes decrease the atmospheric He-abundance
well below solar values in both cases. In fact, we find no difference in surface helium abundance between
$\omega$\,Cen and the other clusters, and $\log{\mathrm (N_{He}/N_H)}\leq -$1.5 for all the stars. With
atmospheres of very similar chemical composition, their SED should not be very different, and even
the known UV-enhanced flux of these stars \citep{Grundahl99} should have the same effects irrespective of the
primordial helium content.

%%%%%%%%%%%%%%%%%%%%%%%%%%%%%%%%%%%%%%%%%
%%%%%%%%%%%%%%%%%%%%%%%%%%%%%%%%%%%%%%%%%

\section{CONCLUSIONS}
\label{s_conclusions}

Blue HB stars in $\omega$\,Cen show lower gravities with respect to both canonical models and analogous stars in
other GCs, but their stellar masses are underestimated, and their visual absolute magnitudes are very similar to
those of the comparison clusters. Neither the low gravities nor the other parameters involved in the calculation
can explain the too low masses. We can firmly conclude that these results reveal an intrinsic difference between
the blue HB stars in $\omega$\,Cen and their analogs in other GCs, but its interpretation is not straightforward.
The lower gravities follow the expectations for He-rich stars, but the magnitudes and masses do not.

\acknowledgments
C.M.B. acknowledges support from the Chilean projects {\sl Centro de Astrof\'\i sica} FONDAP No. 15010003
and the Chilean Centro de Excelencia en Astrof\'\i sica y Tecnolog\'\i as Afines (CATA) BASAL PFB/06.
G.P. acknowledges support by MIUR under the program PRIN2007 (prot.\ 20075TP5K9), and PRIN-INAF 2009.
The authors are grateful to the anonymous referee for his/her fruitful report.


\begin{thebibliography}{}
\bibitem[Alonso et al.(1999)]{Alonso99}
Alonso, A., Arribas, S., \& Mart\'inez-Roger, C. 1999, A\&AS, 140, 261
\bibitem[Bedin et al.(2004)]{Bedin04}
Bedin, L. R., Piotto, G., Anderson, J., Cassisi, S., King, I. R., Momany, Y., \& Carraro, G. 2004, ApJ, 605, L125
\bibitem[Behr et al.(1999)]{Behr99}
Behr, B. B., Cohen, J. G., McCarthy, J. K., \& Djorgovski, S. G. 1999, ApJ, 517, L135
\bibitem[Behr(2003)]{Behr03}
Behr, B. B. 2003, ApJS, 149, 67
\bibitem[Bellini et al.(2009)]{Bellini09}
Bellini, A., et al. 2009, A\&A, 493, 959
\bibitem[Bellini et al.(2010)]{Bellini10}
Bellini, A., Bedin, L.~R., Piotto, G., Milone, A.~P., Marino, A.~F., \& Villanova, S. 2010, AJ, 140, 631
\bibitem[Bergeron et al.(1992)]{Bergeron92}
Bergeron, P., Saffer, R. A., \& Liebert, J. 1992, ApJ, 394, 228
\bibitem[Caloi \& D'Antona(2007)]{Caloi07}
Caloi, V, \& D'Antona, F. 2007, A\&A, 463, 949
\bibitem[Cassisi et al.(1999)]{Cassisi99}
Cassisi, S., Castellani, V., degl'Innocenti, S., Salaris, M., \& Weiss, A. 1999, A\&AS, 134, 103
\bibitem[Castellani et al.(2007)]{Castellani07}
Castellani, V., et al. 2007, ApJ, 663, 1021
\bibitem[Catelan et al.(2009)]{Catelan09}
Catelan, M., Grundahl, F., Sweigart, A. V., Valcarce, A. A. R., \& Cort\'es, C. 2009, ApJ, 695, L97
\bibitem[Catelan(2009)]{Catelan09b}
Catelan, C. 2009, Ap\&SS, 320, 261
\bibitem[D'Antona et al.(2002)]{DAntona02}
D'Antona, F., Caloi, V., Montalb\'an, J., Ventura, P., \& Gratton, R. 2002, A\&A, 395, 69
\bibitem[D'Antona \& Caloi(2004)]{DAntona04}
D'Antona, F., \& Caloi, V. 2004, ApJ, 611, 871
\bibitem[Flower(1996)]{Flower96}
Flower, P. J. 1996, ApJ, 469, 355
\bibitem[Grundahl et al.(1999)]{Grundahl99}
Grundahl, F., Catelan, M., Landsman, W.~B., Stetson, P.~B., \& Andersen, M.~I. 1999, ApJ, 524, 242
\bibitem[Hamuy et al.(1992)]{Hamuy92}
Hamuy, M., Walker, A. R., Suntzeff, N. B., Gigoux, P., Heathcote, S. R., \& Phillips, M. M. 1992, PASP, 104, 533
\bibitem[Harris(1996)]{Harris96}
Harris, W. E. 1996, AJ, 112, 1487
\bibitem[Kurucz(1993)]{Kurucz93}
Kurucz, R. 1993, ATLAS9 Stellar Atmosphere Programs and 2 km/s grid. Kurucz CD-ROM No. 13.
Cambridge, Mass.: Smithsonian Astrophysical Observatory, 1993., 13
\bibitem[LeBlanc et al.(2010)]{Leblanc10}
LeBlanc, F., Hui-Bon-Hoa, A., \& Khalack, V. R. 2010, MNRAS, 409, 1606
\bibitem[Moehler et al.(2011)]{Moehler11}
Moehler, S., Dreizler, S., Lanz, T., Bono, G., Sweigart, A. V., Calamida, A., \& Nonino, M. 2011, A\&A, 526, 136
\bibitem[Moehler et al.(2000)]{Moehler00}
Moehler, S., Landsman, W. B., \& Dorman, B. 2000, A\&A, 361, 937
\bibitem[Moehler et al.(2003)]{Moehler03}
Moehler, S., Landsman, W. B., Sweigart, A. V., \& Grundahl, F. 2003, A\&A, 405, 135
\bibitem[Moehler et al.(1999)]{Moehler99}
Moehler, S., Sweigart, A. V., \& Catelan, M. 1999, A\&A, 351, 519
\bibitem[Moehler et al.(2000)]{Moehler00}
Moehler, S., Sweigart, A. V., Landsman, W. B., \& Heber, U. 2000, A\&A, 360, 120
\bibitem[Moehler \& Sweigart(2006)]{Moehler06}
Moehler, S., \& Sweigart, A. V. 2006, A\&A, 455, 943
\bibitem[Momany et al.(2004)]{Momany04}
Momany, Y., Bedin, L. R., Cassisi, S., Piotto, G., Ortolani, S., Recio-Blanco, A., De Angeli, F., \& Castelli, F. 2004,
A\&A, 420, 605
\bibitem[Moni Bidin et al.(2007)]{Moni07}
Moni Bidin, C., Moehler, S., Piotto, G., Momany, Y., \& Recio-Blanco, A. 2007, A\&A, 474, 505
\bibitem[Moni Bidin et al.(2009)]{Moni09}
Moni Bidin, C., Moehler, S., Piotto, G., Momany, Y., \& Recio-Blanco, A. 2009, A\&A, 498, 737
\bibitem[Napiwotzki et al.(1999)]{Napiwotzki99}
Napiwotzki, R., Green, P. J., \& Saffer, R. A. 1999, ApJ, 517, 399
\bibitem[Norris(2004)]{Norris04}
Norris, J. 2004, ApJ, 612, L25
\bibitem[Pace et al.(2006)]{Pace06}
Pace, G., Recio-Blanco, A., Piotto, G., \& Momany, Y. 2006, A\&A, 452, 493
\bibitem[Pancino et al.(2011)]{Pancino11}
Pancino, E., Mucciarelli, A., Sbordone, L., Bellazzini, M., Pasquini, L., Monaco, L., \& Ferraro, F.~R.
2011, A\&A, 527, A18
\bibitem[Piotto et al.(2005)]{Piotto05}
Piotto, G., et al. 2005, ApJ, 621, 777
\bibitem[Piotto et al.(2007)]{Piotto07}
Piotto, G., et al. 2007, ApJ, 661, L53
\bibitem[Rich et al.(1997)]{Rich97}
Rich, R. M., et al. 1997, ApJ, 484, L25
\bibitem[Saffer et al.(1994)]{Saffer94}
Saffer, R. A., Bergeron, P., Koester, D., \& Liebert, J. 1994, ApJ, 432, 351
\bibitem[Sandage \& Wildey(1967)]{Sandage67}
Sandage, A. \& Wildey, R. 1967, ApJ, 150, 469
\bibitem[Sollima et al.(2005)]{Sollima05}
Sollima, A., Pancino, E., Ferraro, F. R., Bellazzini, M., Straniero, O., \& Pasquini, L. 2005, ApJ, 634, 332
\bibitem[Sweigart(1987)]{Sweigart87}
Sweigart, A. V. 1987, ApJS, 65, 95
\bibitem[Sweigart(1997)]{Sweigart97}
Sweigart, A. V. 1997, ApJ, 474, L23
\bibitem[Tonry \& Davis(1979)]{Tonry79}
Tonry, J. \& Davis, M. 1979, AJ, 84, 1511
\bibitem[van de Ven et al.(2006)]{Vandeven06}
van de Ven, G., van den Bosch, R. C. E., Verolme, E. K., \& de Zeeuw, P. T. 2006, A\&A, 445, 513

\end{thebibliography}
\end{document}